\documentclass[%
 reprint,
 amsmath,amssymb,
 aps,
]{revtex4-1}

\usepackage{graphicx}
\usepackage{dcolumn}
\usepackage{bm}
\usepackage{soul}
\usepackage{hyperref}

\usepackage{xcolor}
\begin{document}


\title{Theory of Vacuum Texture: Blackbody Radiation, Uncertainty Principle and Quantum Statistics}
\author{Yoko Suzuki}\email{ysuzuki@newmexicoconsortium.org}
\affiliation{
 New Mexico Consortium, Los Alamos, NM 87544, USA
}
\author{Kevin M Mertes}
\email{kmmertes@lanl.gov}
\affiliation{
Los Alamos National Laboratory, Los Alamos, NM 87545, USA
}

\date{\today}

\begin{abstract}
Previously, we presented a new interpretation of quantum mechanics that revealed it is indeed possible to have a local hidden variable that is consistent with Bell's inequality experiments.  In that article we suggested that the local hidden variable is associated with vacuum fluctuations.  In this article we expound upon that notion by introducing the Theory of Vacuum Texture (TVT).  Here we show that replacing the highly restrictive assumptions of the quantization of energy levels in a system with the simpler, less restrictive postulate that there exists a threshold in order for energy to be released. With this new postulate, the models of blackbody radiation is shown to be consistent with the experiments. We also show, that the threshold condition contributes to a localized vacuum energy which leads us to conclude that the uncertainty principle is a statistical effect. These conditions also naturally leads to the prediction that massive particles transition to an ordered state at low temperatures.  In addition, we show that thermodynamic laws must be modified to include two heat baths with temperatures: $T$ for dissipative energy levels and $T_{V}$ ($\gg T$) for localized vacuum energy.  In total, we show that our threshold postulate agrees with experimental observations of blackbody radiation, the uncertainty principle and quantum statistics without the need of the invoking quantum weirdness.
\begin{description}
\item[PACS numbers]
03.65.-w, 03.65.Ta, 03.65.Ud
\end{description}
\end{abstract}

\pacs{Valid PACS appear here}
\maketitle

\section{\label{sec:Introduction}Introduction}
We have shown that a local variable theory \cite{Yoko1} can be consistent with Bell’s inequality experiments and suggested an experimental method to prove the existence of vacuum texture. In this article, we will introduce the origin of vacuum texture. The theory of vacuum texture (TVT) is a microscopic theory of the vacuum which is a deterministic theory with local variables. We will propose one assumption about the microscopic properties of the vacuum and compute statistical quantities to see if they are consistent with experimental observations.

The quantization of energy levels was first suggested in order to avoid the ultraviolet catastrophe for blackbody radiation \cite{Plank}.  Measurements clearly showed the absence of high frequency radiation in the cavity. A mathematical formulation, which fit the data well, was evaluated and deemed sufficient.  This lead to the development of Quantum Mechanics and the interpretation of that formulation has been debated since. 

We will establish a new formulation which fits the data as well as, if not better than, Planck's quantization theory. TVT has been formulated based mainly on the self-similarity principle and consistency with all aspects of experimental observations. This formulation, unlike quantum mechanics, is intuitive, realistic and self-consistent. It is straight forward, and there is nothing ``spooky'' about it. It also might explain other aspects of nature such as dark matter and the history of the universe \cite{Yoko2}.

\section{\label{sec:Postulates}Ultraviolet Catastrophy}

When Rayleigh and Jeans \cite{RayleighJeans} combined the classical theory of radiation with the theory of thermodynamics, they revealed an inconsistency within the then known laws of physics that has been termed the ``ultraviolet catastrophe''. It is well understood that the large discrepancy between theoretical prediction and experimental observation occurred because it was assumed the probability of mode excitation, $e^{-E/k_B T}$, did not depend on the frequency.  Planck was able to avoid the ultraviolet catastrophe by assuming that only discrete energy levels for the radiation in the cavity were allowed.  This ultimately paved the path for the development of Quantum Mechanics.  

\begin{figure}[h]
\centering
\includegraphics[trim={0cm  0.8cm  0cm 0cm},clip,width=0.48\textwidth]{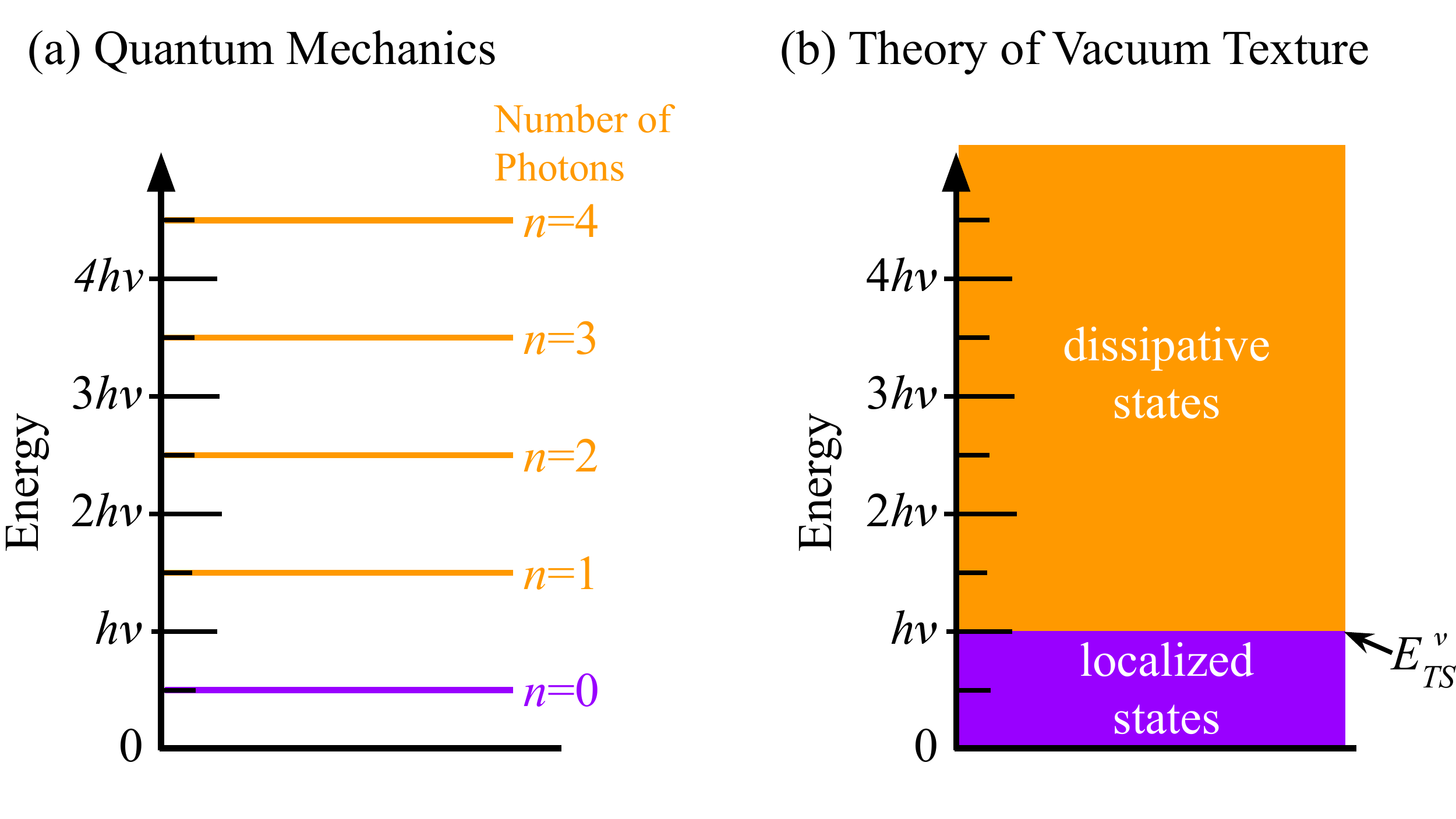}
\caption{The allowed energy states of the vacuum and electromagnetic fields associated with frequency, $\nu$ are shown for quantum mechanics in (a) and TVT in (b). Conventionally, the orange part (dissipative states) are defined as electromagnetic fields (or photons), in which, the energy travels with speed of light. The purple part (localized states) are defined as the vacuum, in which, there is no light-speed spatial energy transfer. Any state is allowed for TVT in (b). However, it has a threshold value, $E_{TS}^\nu=h \nu$ which separates the localized energy states and the dissipative energy states for each $\nu$.}
\label{fig:States}
\end{figure}

However, we will show that the essential ingredient that allowed Planck to avoid the ultraviolet catastrophe was not the quantization of energy levels, but rather the absence of radiation at frequencies, $\nu$, when $E<h\nu$. The comparison between Planck's principle of the quantization of energy levels and TVT is shown in Fig. \ref{fig:States}. The postulates of TVT is much less restrictive than the ones of quantum mechanics. It relaxes all the constraints of quantum mechanics except for the existence of a threshold energy. 

\section{\label{sec:Postulates}New Postulate of the vacuum}
In contrast to quantum mechanics, we assume that all energy levels above and below the threshold are continuously allowed, but below the threshold, the energy will not be dissipated in space with the speed of light. We call these dissipative and localized states, respectively. If the experimental findings can be explained by the postulate in Fig. \ref{fig:States} (b), there is no need to have such restrictive postulates as in Fig. \ref{fig:States} (a).  
However, prohibiting the creation of low energy states is unnatural especially when we already know the vacuum is filled with such energy due to vacuum fluctuations. Thus, we will only prohibit low energy states from dissipating through space perhaps due to some intrinsic property of the vacuum at low energy with relatively high $\nu$. In other words, all energy levels are allowed; however, there is a threshold condition for energy to propagate at the speed of light for each $\nu$. The key new assumption we propose for the description of the vacuum is:
\begin{itemize}
  \item[] For each $\nu$, space can not transport energy at the speed of light when $E^\nu_{TS}<h \nu$.
\end{itemize}

The remainder of the article will focus on the consequences of assuming the above definition.

\section{\label{sec:blackbody radiation}blackbody radiation}

The spectral radiance, $B_\nu$ of blackbody radiation for frequency $\nu$ at temperature $T$ is described by Planck's law:
\begin{equation}
B_\nu^{qm}(\nu, T)=\frac{2\nu^2}{c^2}\frac{h\nu}{e^{\frac{h\nu}{k_BT}}-1}=\frac{2\nu^2}{c^2}h\nu f_{BE}(h\nu/k_BT)
\label{eq:Bqm}
\end{equation}
where $k_B$ is the Boltzmann constant, $h$ is the Planck constant, and $c$ is the speed of light. This was derived from the interpretation of photons as indistinguishable particles in the Bose-Einstein distribution \cite{Bose, Einstein},
\begin{equation}
f_{BE}(x)=\frac{1}{e^{x}-1}.
\label{eq:FBE}
\end{equation}

\begin{figure}[h]
 \includegraphics[trim={1.5cm  5.5cm  2cm 0.5cm}, clip, width=0.48\textwidth]{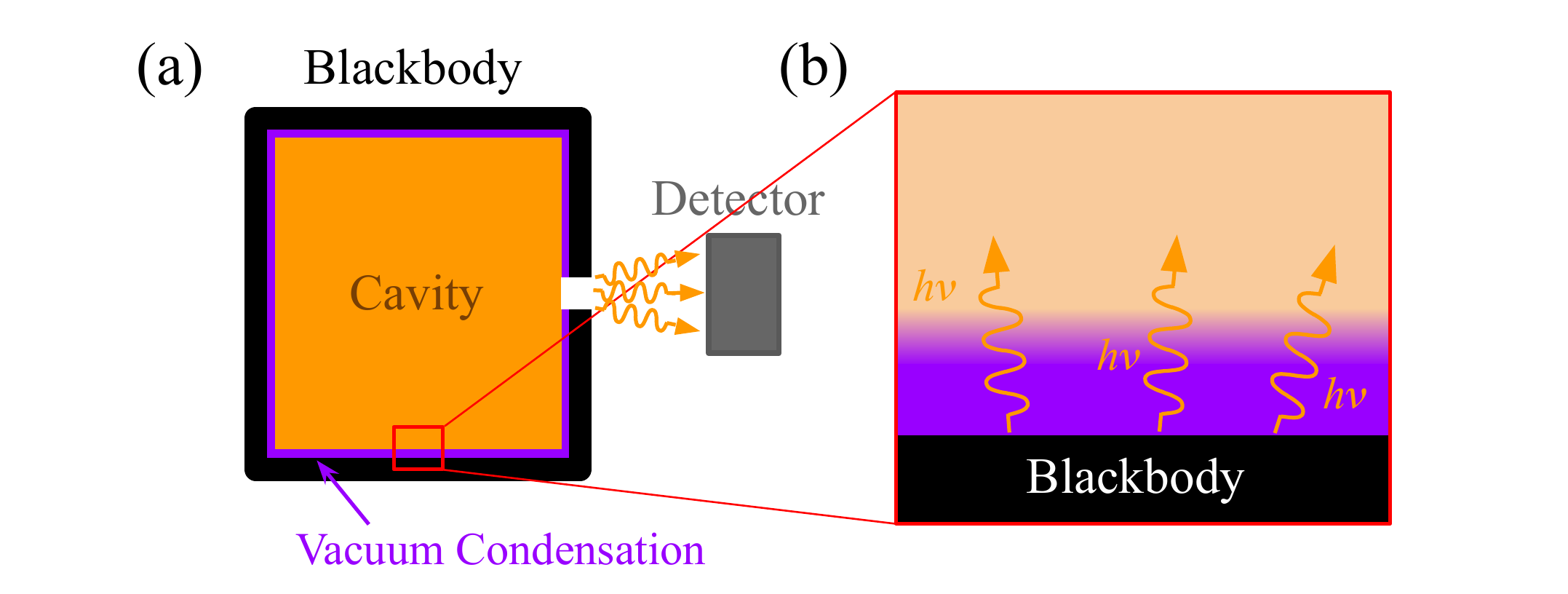}
\caption{\textbf{(a)} The schematic of a \textit{gedankenexperiment} of blackbody radiation is shown. \textbf{(b)} The microscopic schematic of blackbody surface with vacuum condensation (purple part also in Fig. \ref{fig:States} (b)) is shown. In order to be consistent with the experiments, the vacuum energy needs to be assumed to be condensed on the surface of the blackbody. If it drifts away, it would eventually fill the whole cavity and would reach the detector. Therefore, it would cause the ultraviolet catastrophe.}
\label{fig:BBschematic}
\end{figure}

In TVT, all the energy levels above the threshold are allowed. However, due to the threshold condition, the energy below $h\nu$ do not leave the vicinity of the radiation emitter. Fig. \ref{fig:BBschematic} (a) shows that even if the blackbody is a perfectly-smooth spectral emitter, the purple part (localized energy) would never reach the detector. The only energy which is larger than $h\nu$ would propagate to the detector (see Fig. \ref{fig:BBschematic} (b)). According to the postulate, as soon as the energy reaches the threshold condition, $E^\nu_{TS}=h\nu$, it dissipates from the emitter with the speed of light. Because of that, mostly $h\nu$ might be emitted. Nevertheless, at this point, all the states above the threshold are used to derive the spectral radiance simply by using Maxwell–Boltzmann statistics.
\begin{multline}
\label{eq:Bvc}
B_\nu^{V}(\nu, T)
=\frac{2\nu^2}{c^2}\frac{\int_{h\nu}^\infty E\,e^{-\frac{E}{k_B T}}dE}{\int_{0}^\infty e^{-\frac{E}{k_B T}}dE}\\
=\frac{2\nu^2}{c^2}(h\nu+k_B T)e^{-\frac{h\nu}{k_BT}},
\end{multline}
where we only integrate the numberator from $h\nu$ to $\infty$.
If the lower limit of the integral in the numerator of Eq. \ref{eq:Bvc} is $0$ instead of $h\nu$, it would be identical to the classical model of Rayleigh \& Jeans, $B_\nu^{cl}(\nu, T)=\frac{2\nu^2}{c^2}k_B T$. This energy in the integral between 0 and $h\nu$ which was subtracted in Eq. \ref{eq:Bvc} from the classical model will be analyzed in more detail in Sec. \ref{sec:UncertaintyPrincipal}.

\begin{figure}[h]
\includegraphics[trim={0cm  0cm  0cm 0cm},clip,width=0.49\textwidth]{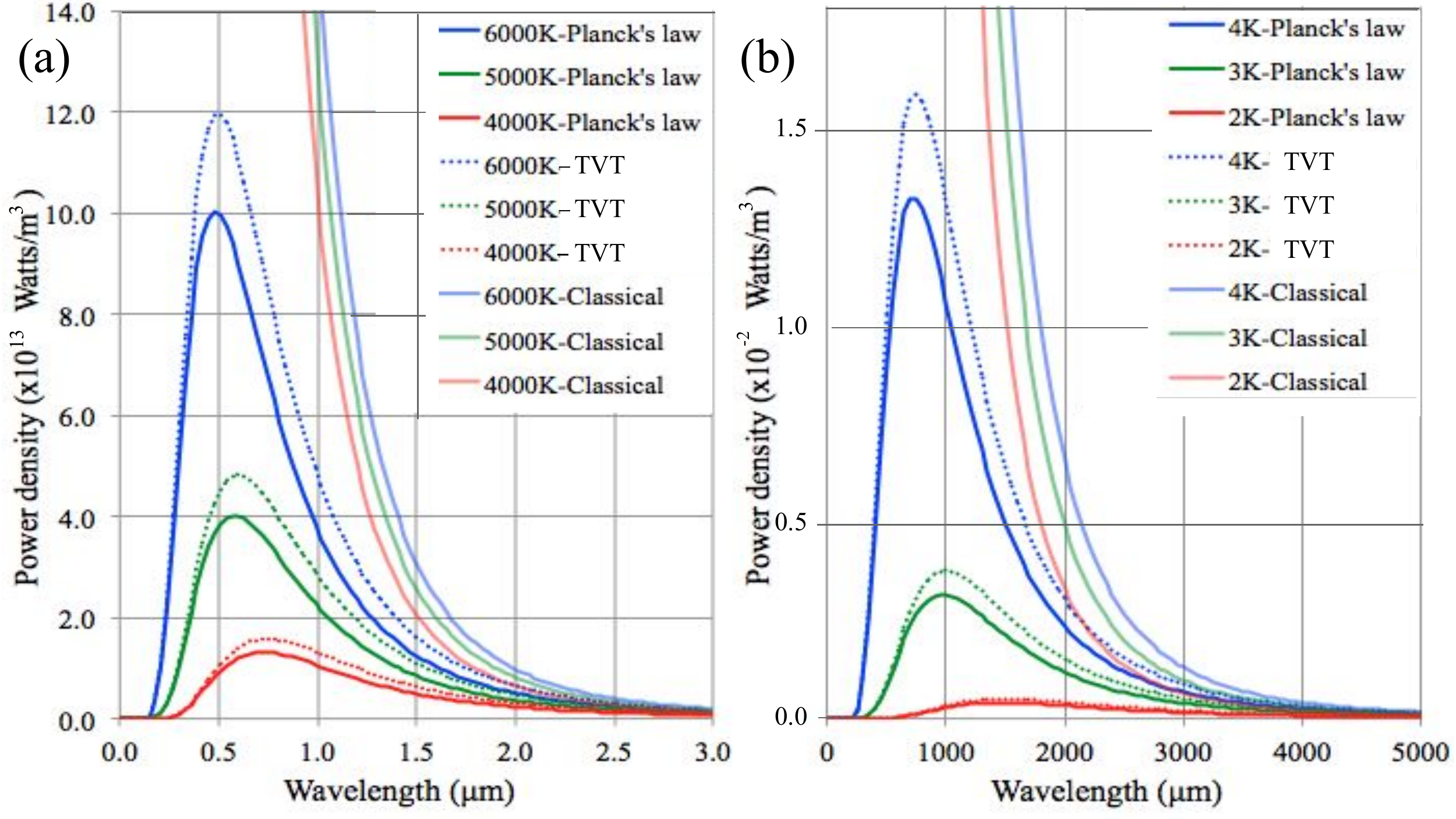}
\caption{The radiated power density per wave length, $S_\lambda(\lambda, T)$ is plotted for 6000 K, 5000 K \& 4000 K in (a) and 4 K, 3 K \& 2 K in (b) of Planck's law, TVT and the classical model of Rayleigh \& Jeans .}
\label{fig:BB}
\end{figure}

The radiated power density per wavelength is, 
\begin{align}
S_\lambda^{qm}(\lambda, T)&=\frac{2\pi c}{\lambda^4}\frac{hc/\lambda}{e^{hc/\lambda k_B T}-1}\\
S_\lambda^{V}(\lambda, T)&=\frac{2\pi c}{\lambda^4}(\frac{hc}{\lambda}+k_B T)e^{-hc/\lambda k_B T}\\
S_\lambda^{cl}(\lambda, T)&=\frac{2\pi c}{\lambda^4}k_B T
\end{align}
for quantum mechanics, TVT and the classical model of Rayleigh \& Jeans, respectively.  In Fig. \ref{fig:BB}, we plot them in order to compare with experiment. It is well known that Plank's law of $S_\lambda^{qm}(\lambda, T)$ fits the experimental finding very well at 5777 K for the sun and at 2.7 K for the cosmic microwave background (CMB). Thus, TVT gives similar values to plank's law. Clearly, the ultraviolet catastrophe is successfully avoided with TVT, with the same degree of accuracy as quantum mechanics.

One of the accolades placed on Planck's quantization principle was the derivation of the  Stefan-Boltzmann law \cite{Stefan, Boltzmann}. Under the Planck model, the energy density of blackbody radiation is:
\begin{equation}
U=\int_0^\infty \frac{4\pi}{c} B_\nu^{qm}(\nu, T)d\nu=\frac{4}{c}\sigma T^4
\label{eq:Uqm}
\end{equation}
where $\sigma=2\pi^5 k_B^4/15 h^3 c^2$ is the Stefan-Boltzmann constant. 

If we use $B_\nu^{V}(\nu, T)$ from TVT in Eq. \ref{eq:Bvc} instead,
\begin{multline}
U=\int_0^\infty\frac{4\pi}{c} B_\nu^{V}(\nu, T)d\nu\\
=\frac{8\pi k_B^4}{h^3 c^3} T^4\int_0^\infty u^2 (u+1)e^{-u}du
=\frac{4}{c}\sigma^{V} T^4
\end{multline}
where $u=h\nu/k_B T$ and $\sigma^{V}=16\pi k_B^4/h^3 c^2\approx 1.23\sigma$.
Thus we see that the  Stefan-Boltzmann law also can be successfully derived from TVT albeit with a slightly larger constant and a more reasonable numerical factor \cite{Greenberger}.

In addition, Plank's law predicts Wien's displacement law $\lambda_{max}=b/T$ where $b=2.90\times10^{-3}$ mK \cite{Wien}. If we use TVT in Eq. \ref{eq:Bvc} instead,
\begin{multline}
\frac{\partial S_\lambda^{V}(\lambda, T)}{\partial \lambda}=2\pi c (\frac{h^2 c^2}{k_B T \lambda^7}-\frac{4hc}{\lambda^6}-\frac{4k_B T}{\lambda^5}) e^{-\frac{hc}{k_B T \lambda}}=0
\end{multline}
$\lambda^{V}_{max}=b^{V}/T$ where $b^{V}=2.98\times10^{-3}$ mK $\approx 1.03b$. TVT also predicts Wien's displacement law successfully with a slightly larger constant, and again the numerical factor is obtained in more effortless way.

\section{\label{sec:UncertaintyPrincipal}Uncertainty Principal}
In quantum mechanics, the uncertainty principle, (i.e., the existence of zero-point energy) is essential to  accurately describe the behavior of electrons in atoms. Without the existence of the electronic ground state, an electron would fall into the nucleus quite quickly. In this section, the localized states (purple part in Fig. \ref{fig:States} (b)) will be analyzed by introducing the vacuum energy, $\epsilon_0^{V}(\nu)$ for frequency $\nu$. In quantum mechanics,
\begin{equation}
\epsilon_0^{qm}(\nu)=\frac{h\nu}{2}
\end{equation}
which is the energy for $n=0$ in Fig. \ref{fig:States} (a).

\begin{figure}[h]
\includegraphics[trim={0cm  0cm  0cm 0cm},clip,width=0.48\textwidth]{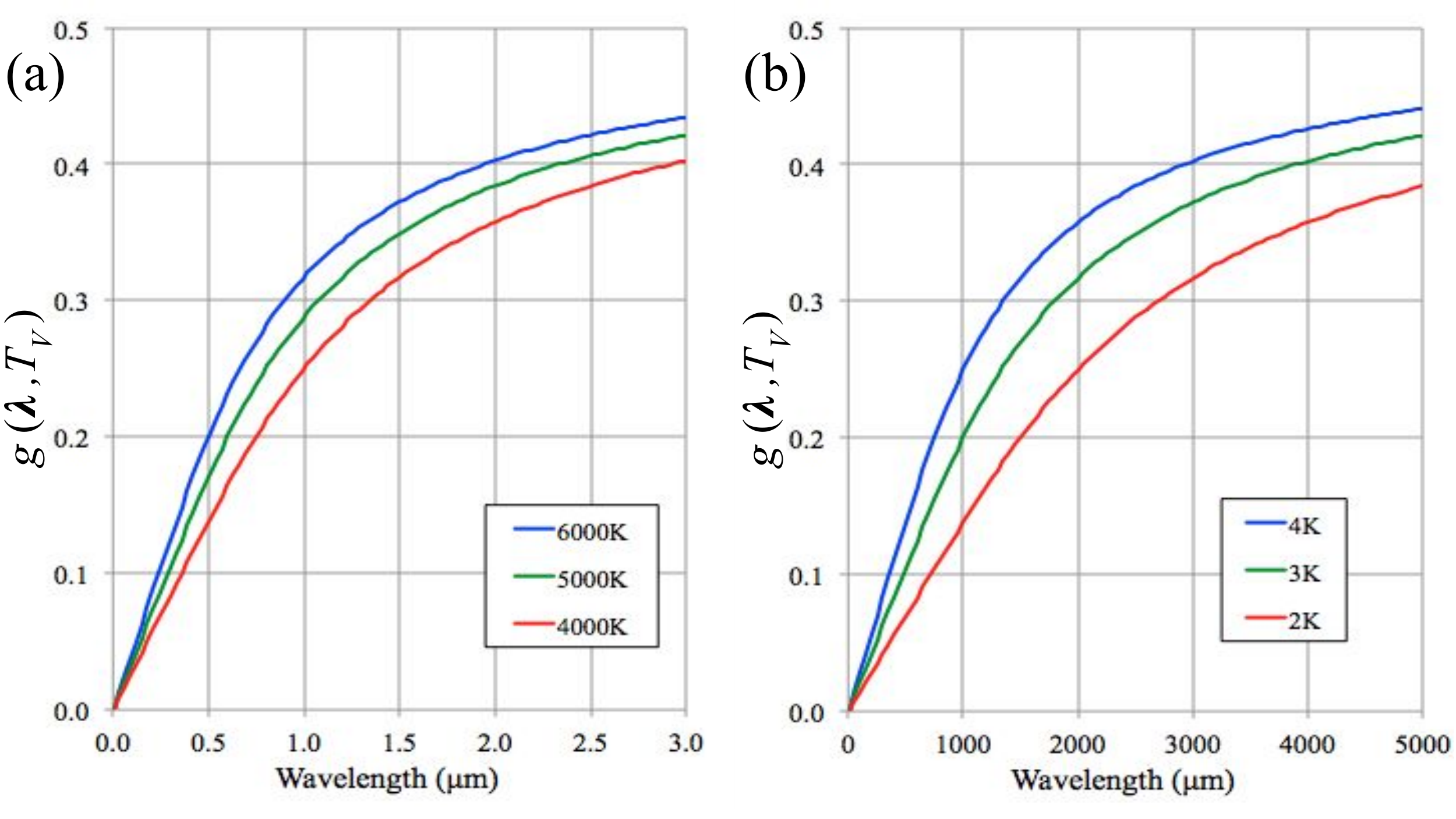}
\caption{Eq. \ref{eq:gvc} is plotted for 6000 K, 5000 K \& 4000 K in (a) and 4 K, 3 K \& 2 K in (b), as a function of wavelength similar to Fig. \ref{fig:BB}}
\label{fig:g}
\end{figure}

For TVT, the average vacuum energy for \textit{one} localized state due to vacuum temperature, $T_V$ is defined by taking an average value of the localized states (purple part in Fig. \ref{fig:States} (b)) simply using the Boltzmann distribution:
\begin{multline}
\label{eq:Evc}
\epsilon_0^{V}(\nu, T_V)
=\frac{\int_{0}^{h\nu} E\,e^{-\frac{E}{k_B T_V}}dE}{\int_{0}^{h\nu} e^{-\frac{E}{k_B T_V}}dE}
=k_B T_V-\frac{h\nu}{e^{\frac{h\nu}{k_B T_V}}-1}\\
=g(\nu, T_V)h\nu
\end{multline}
where
\begin{equation}
\label{eq:gvc}
g(\nu, T_V)
=\frac{k_B T_V}{h\nu}-\frac{1}{e^{\frac{h\nu}{k_B T_V}}-1}.
\end{equation}
$g(\nu, T_V)\rightarrow 1/2$ when $h\nu \ll k_B T_V$. It is plotted in Fig. \ref{fig:g} as a function of wavelength.

Now, we would like to review the processes in a \textit{gedankenexperiment} involving blackbody radiation. In Fig. \ref{fig:BBschematic} (a), before the experiment, the blackbody and the cavity are all cold. As the heater heats up the blackbody, it emits radiation into the cavity. It emits more radiation than it absorbs until it reaches the thermal equilibrium with the cavity at temperature $T$. After that, the intensities of emission and absorption are equal, and the heater is turned off. In TVT, the blackbody also fills the localized vacuum states (purple states in Fig. \ref{fig:States} (b)); on the interior surface, the blackbody emitter simultaneously tends toward the saturation value, $\epsilon_0^{V}(\nu, T)$ as shown in Eq. \ref{eq:Evc}. However, this energy will not propagate with speed of light according to our postulate. Instead, it will equilibrate by drifting around the interior surface of the emitter, and the emitter will keep filling the localized vacuum states until it reaches thermal equilibrium over the entire interior surface of the emitter at temperature $T$. After that, the intensities of emission and absorption are equal, and the heater is turned off.

The question is how much vacuum energy would drift away in space and ooze out of the blackbody walls? In other words, what is the thickness of purple color in Fig. \ref{fig:BBschematic}? If it oozes out far enough to fill the entire cavity and toward the detector, the ultraviolet catastrophe would happen. Therefore, the assumption that the vacuum energy condenses rather than scatters away is necessary in order to be consistent with experimental observation.

\begin{figure}[h]
\includegraphics[trim={1cm  3cm  1cm 0cm},clip,width=0.4\textwidth]{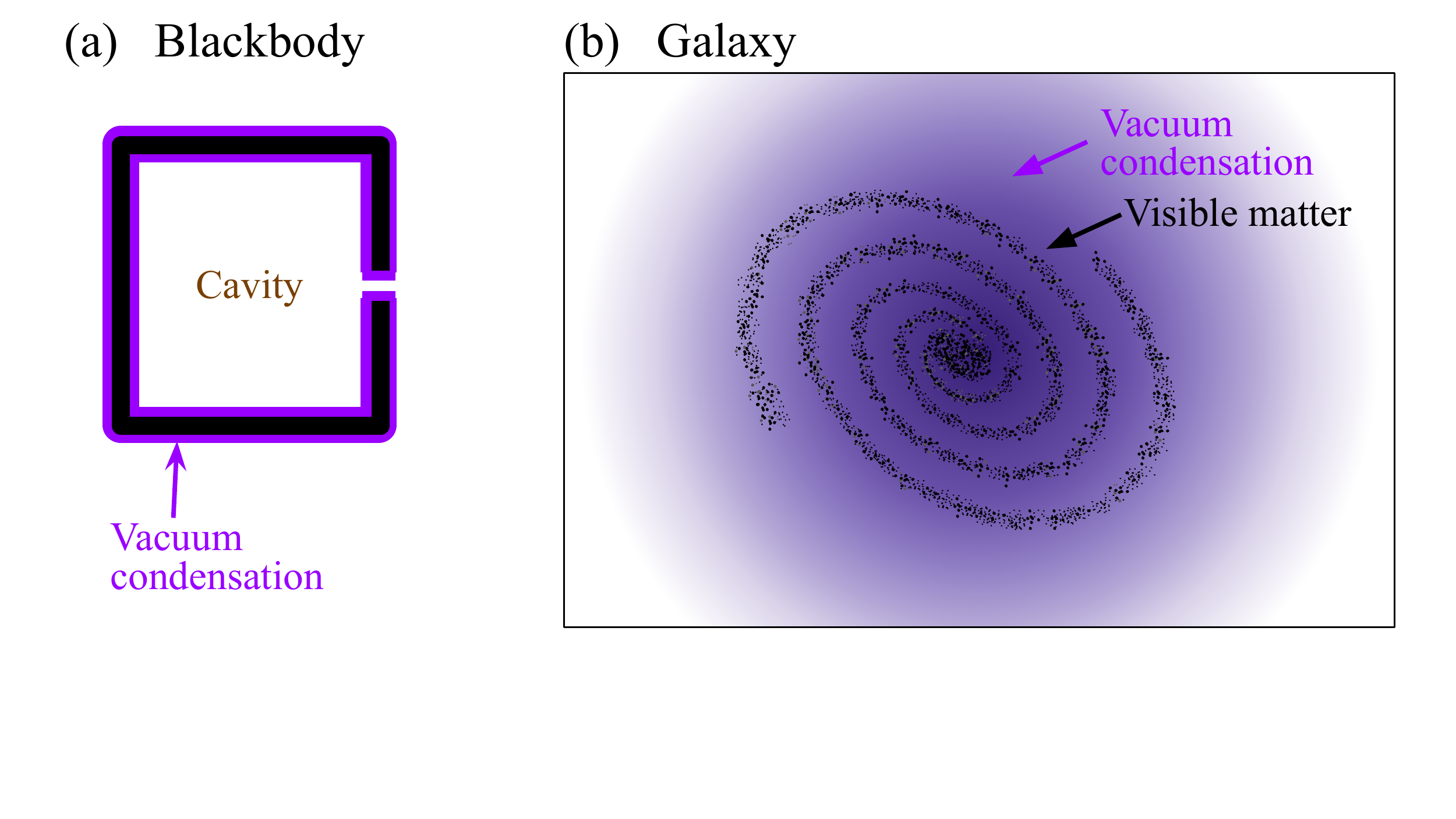}
\caption{\textbf{(a)} The schematic of a \textit{gedankenexperiment} of blackbody radiation when it has cooled down completely after the initial heating in Fig. \ref{fig:BBschematic}. While the energy in dissipative states (orange area) have dissipated away, the localized energy states (purple area) which are below the threshold energy values are condensed and oscillating indefinitely because they do not have any mechanism to dissipate nor disappear. Therefore, the blackbody radiation experiment would depend on the thermal history of the blackbody. \textbf{(b)} The big bang theory assumes that the whole universe was once at high temperature before it expanded. Therefore, we surmise that visible matter (e.g., large structures like galaxies to fundamental particles like electrons) are surrounded by vacuum condensation.}
\label{fig:Galaxy}
\end{figure}

On the other hand, thermal history of the blackbody is very important (see caption of Fig. \ref{fig:Galaxy} for a detailed discussion). In other words, the vacuum exhibits thermal hysteresis. In reality, overwhelming evidence shows that the vacuum was heated to a very high temperature in the beginning of the universe. The ultraviolet catastrophe seemed to really happen at the beginning of the universe. When the universe was small, the temperature was higher. As it expanded, the vacuum energy which was created at $T_V\sim 10^{12}-10^{32}$ K during the initial cooling and expansion of the universe. The amount and temperature of vacuum condensation will be estimated with astronomical observations and is discussed in the article \cite{Yoko2}. Please note that the vacuum condensation could be quite inhomogeneous in space. However, in this article we will assume $h\nu \ll k_B T_V$ for each $\nu$. The vacuum interaction at high $\nu$ such in the quantum chromodynamics vacuum will be discussed in another article.

\begin{figure}[h]
\includegraphics[trim={2cm  6.5cm  2.5cm 0.5cm},clip,width=0.49
\textwidth]{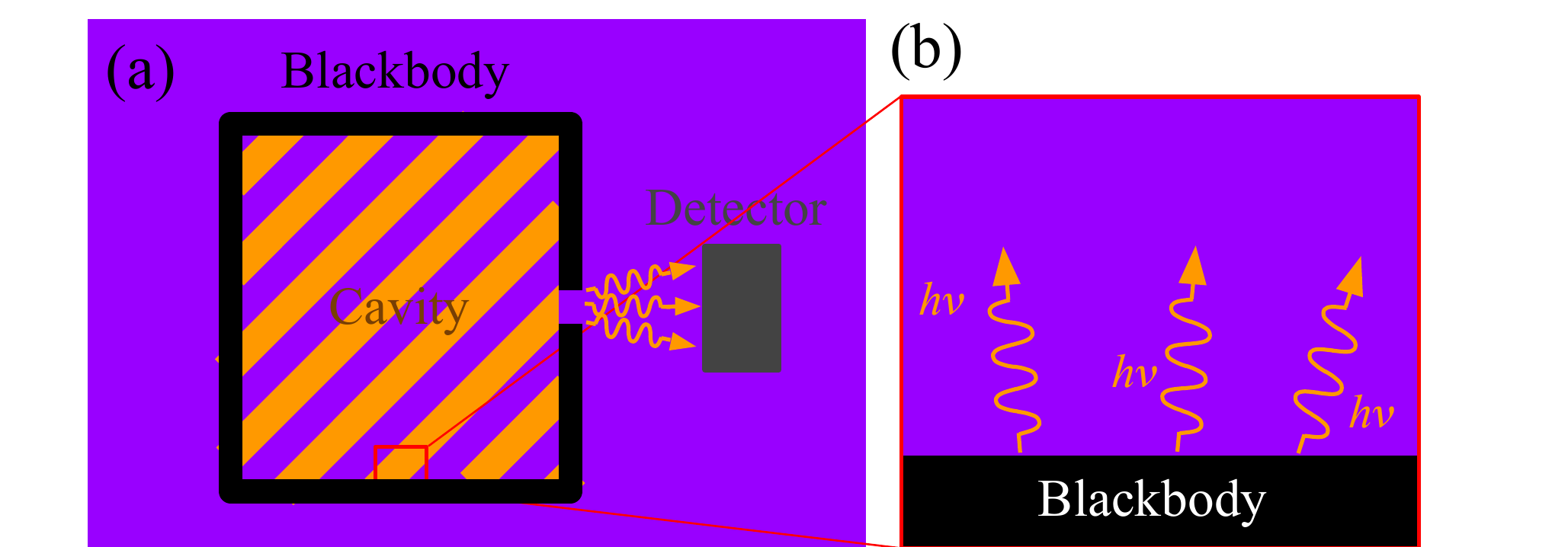}
\caption{\textbf{(a)} The schematic of a \textit{gedankenexperiment} of blackbody radiation in the \textit{filled} vacuum is shown while Fig. \ref{fig:BBschematic} is in the \textit{empty} vacuum. \textbf{(b)} The microscopic schematic of blackbody surface with filled vacuum is shown. }
\label{fig:BBschematicF2}
\end{figure}

Under the condition of $h\nu \ll k_B T_V$ for each $\nu$ which is involved in certain physical phenomena or experiments, $g(\nu, T_V)\rightarrow 1/2$ and $\epsilon_0^{V}(\nu)=\epsilon_\nu^{qm}(\nu)=h\nu/2$. Consequently, TVT is consistent with the experimental findings that have been attributed to the uncertainty principle in Quantum Mechanics. In TVT, uncertainty arises from a statistical effect via vacuum interaction which is not completely random nor uncertain. The interaction between vacuum condensation and visible matter provides us with the means to establish the existence of a hidden, local variable that we alluded to in Ref. \cite{Yoko1} and it is deterministic. In TVT, the vacuum energy and its uncertainty-principle-like effect are \textit{not} the fundamental property in the universe. The condensation of vacuum energy happens to be there because of the way the universe evolved. It is uncertain just because of the statistical effect of so many factors with so many variables. 

\begin{figure}[h]
 \includegraphics[trim={0cm  6.0cm 9cm 0.8cm},clip,width=0.45
 \textwidth]{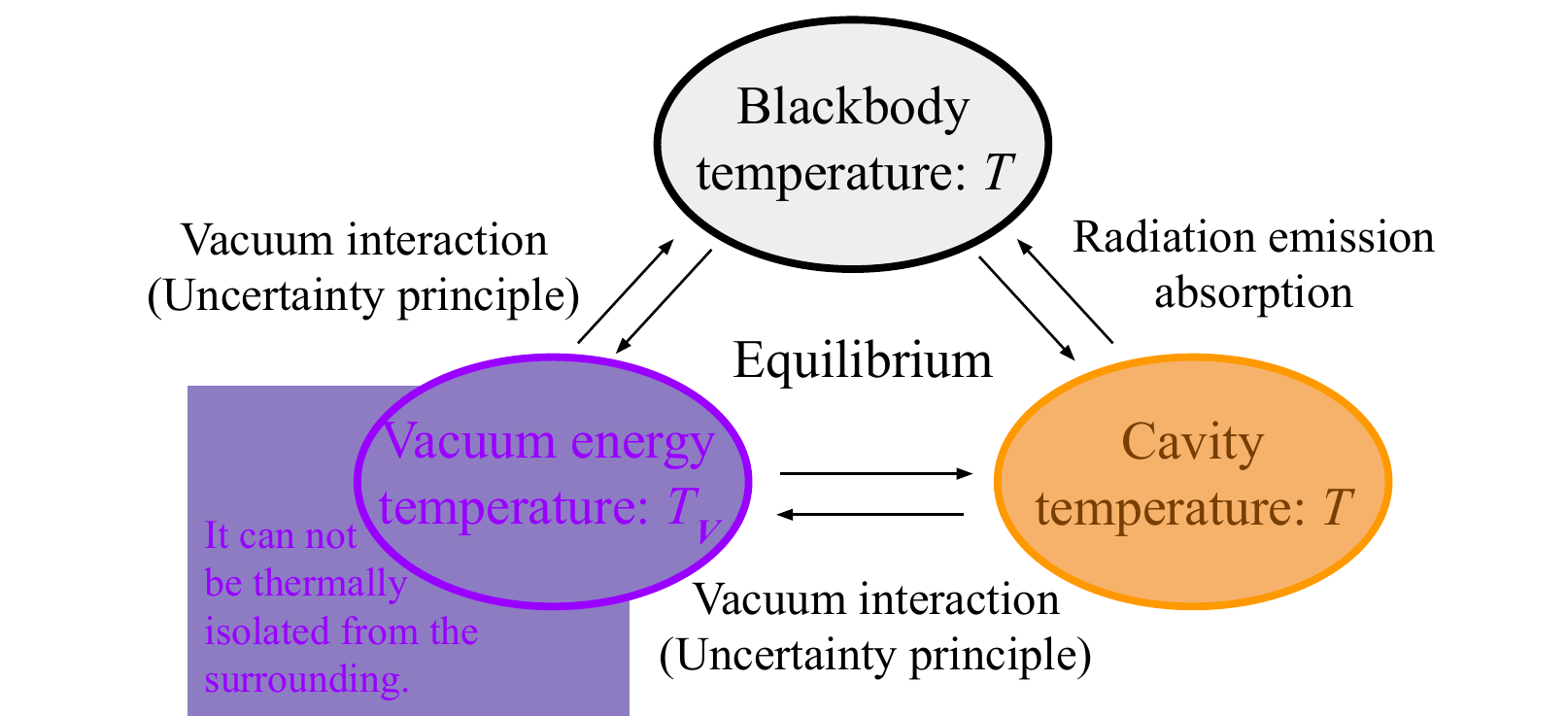}
\caption{The schematic of three-body interaction and two-temperate baths, $T$ \& $T_V$ for a \textit{gedankenexperiment} of blackbody radiation in the filled vacuum condensation at $T_V$ with $T_V\gg T$. The vacuum interaction can result in energy gain or loss for the vacuum in the cavity. However, it is not directly measurable with technology based on our current models. The vacuum in the cavity would quickly equilibrate with the surrounding vacuum at $T_V$.}
\label{fig:BBschematic2}
\end{figure}

Fig. \ref{fig:BBschematicF2} shows the revised schematic for blackbody radiation. The detector is not able to measure the vacuum energy (purple part), rather it only measures the radiation emitted from the cavity (which is the same as shown in Fig. \ref{fig:BBschematic} for the empty vacuum). Also, we need to take into account the fact that before the experiment started the vacuum condensation was already present in the system. The details will be explained in Sec. \ref{sec:blackbody radiation in filled vacuum}. Fig. \ref{fig:BBschematic2} shows that three-body interaction for blackbody radiation with filled vacuum. We assume that the vacuum energy was condensed at very high temperature of $T_V$. Here, it means that the Boltzmann factor, $e^{-E/k_B T_V}\approx 1$ at $0 \leq E \leq h\nu$ for measurements involving $\nu \ll k_B T_V/h$. All matter is directly immersed in the vacuum condensation which is assumed to be highly interactive with other matter. We can safely assume that all other matter have the energy levels at $0 \leq E \leq h\nu$ completely filled as well. In TVT, the interactions between vacuum and matter are called vacuum interactions, and they are locally deterministic and causal. The statistical effects of filled vacuum resemble the behavior of uncertainty principle of quantum mechanics \cite{Boyer}.

\section{\label{sec:blackbody radiation in filled vacuum}blackbody radiation with filled vacuum}

\begin{figure}[h]
\includegraphics[trim={2cm  0.3cm  2cm 0cm},clip,width=0.4\textwidth]{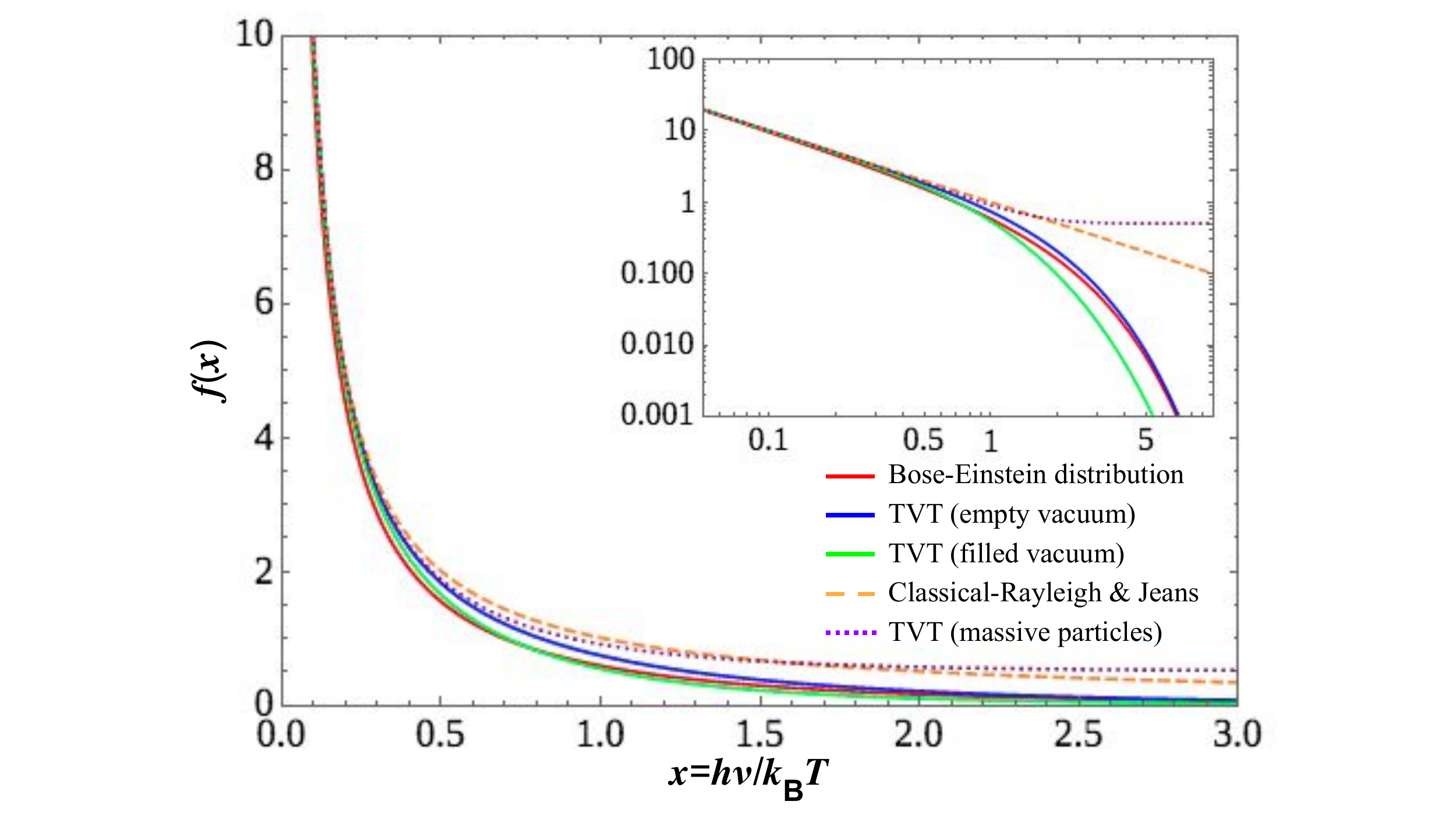}
\caption{The distribution functions are plotted: Bose-Einstein distribution in Eq. \ref{eq:FBE} (red), the distribution function for TVT with empty vacuum in Eq. \ref{eq:Fvc} (blue) and the distribution function for TVT with filled vacuum in Eq. \ref{eq:Ffvc} (green). These three functions have similar behavior. The inset is plotted on a logarithmic scale. The distribution for the classical model of Rayleigh \& Jeans, where $B_\nu^{cl}(\nu, T)=\frac{2\nu^2}{c^2}k_B T=\frac{2\nu^2}{c^2}h\nu\ f_{CL}(h\nu/k_BT)$ and $f_{CL}(x)=1/x$, is plotted in dashed orange, and the distribution for massive particles in Eq. \ref{eq:Fmvc} is also plotted in dotted purple.}
\label{fig:f}
\end{figure}

Theories of thermodynamics will need adjustments to take into account of the existence of vacuum condensation. In this article, we would like to make a few easy quick approximations to see if TVT is a viable theory. For TVT with filled vacuum, the spectral radiance, $B_\nu^{FV}(\nu,T)$ is calculated again simply using the Boltzmann factor with $T_V$ for $0 \leq E \leq h\nu$ and with $T$ for $E > h\nu$. However, the energy at $0 \leq E \leq h\nu$ is not included in $B_\nu^{FV}(\nu,T)$ because that energy will not be measured by the detector.
\begin{multline}
B_\nu^{FV}(\nu,T)
=\lim_{T_V\to\infty}\frac{2\nu^2}{c^2}\frac{\int_{h\nu}^\infty E\,e^{-\frac{E}{k_B T}}dE}{\int_{0}^{h\nu} e^{-\frac{E}{k_B T_V}}dE+\int_{h\nu}^\infty e^{-\frac{E}{k_B T}}dE}\\
=\frac{2\nu^2}{c^2}\frac{h\nu+k_B T}{\frac{h\nu}{k_B T}e^{\frac{h\nu}{k_BT}}+1}=\frac{2\nu^2}{c^2}h\nu\ f_{FV}(h\nu/k_BT)
\label{eq:Bfvc}
\end{multline}
where
\begin{equation}
f_{FV}(x)=\frac{1+\frac{1}{x}}{xe^{x}+1}.
\label{eq:Ffvc}
\end{equation}
$f_{FV}(h\nu/k_BT)$ is the distribution function for TVT with filled vacuum, and it is the average fraction of energy relative to $h\nu$ per mode for the frequency $\nu$ at temperature T. Eq. \ref{eq:Bvc} is the spectral radiance for TVT with empty vacuum, and the distribution function is expressed as
\begin{equation}
f_{V}(x)=(1+\frac{1}{x})e^{-x}.
\label{eq:Fvc}
\end{equation}
where $B_\nu^{V}(\nu,T)=\frac{2\nu^2}{c^2}h\nu\ f_{V}(h\nu/k_BT)$. These distribution functions are plotted in Fig. \ref{fig:f}.

Partially filled vacuum with $T_V=\alpha T$ is expressed as
\begin{multline}
\label{eq:Bpvc}
B_\nu^{PV}(\nu,T,\alpha)
=\frac{2\nu^2}{c^2}\frac{\int_{h\nu}^\infty E\,e^{-\frac{E}{k_B T}}dE}{\int_{0}^{h\nu} e^{-\frac{E}{k_B \alpha T}}dE+\int_{h\nu}^\infty e^{-\frac{E}{k_B T}}dE}\\
=\frac{2\nu^2}{c^2}\frac{e^{-\frac{h\nu}{k_BT}}(h\nu+k_B T)}{\alpha +e^{-\frac{h\nu}{k_BT}}-\alpha e^{-\frac{h\nu}{\alpha k_BT}}}=\frac{2\nu^2}{c^2}h\nu\ f^\alpha_{PV}(h\nu/k_BT)
\end{multline}
where
\begin{equation}
f^\alpha_{PV}(x)=\frac{e^{-x}(1+\frac{1}{x})}{\alpha+e^{-x}-\alpha e^{\frac{-x}{\alpha}}}.
\label{eq:Fpvc}
\end{equation}
 It is easy to see $B_\nu^{PV}(\nu,T,\alpha=1)=B_\nu^{V}(\nu,T)$ and  $\lim_{\alpha \to \infty} B_\nu^{PV}(\nu,T,\alpha)=B_\nu^{FV}(\nu,T)$ \cite{alphaT}.

Stefan-Boltzmann law can be also successfully derived from TVT with filled vacuum in Eq. \ref{eq:Bfvc},
\begin{multline}
U^{FV}=\int_0^\infty \frac{4\pi}{c}B_\nu^{FV}(\nu, T)d\nu\\
=\frac{8\pi k_B^4}{h^3 c^3} T^4\int_0^\infty u^2 \frac{u+1}{ue^{u}+1}du
=\frac{4}{c}\sigma^{FV} T^4
\label{eq:Ufvc}
\end{multline}
where $u=h\nu/k_B T$ and $\sigma^{FV}\approx 0.41\sigma$. We should not worry about the factor, 0.41 difference from Planck's law. The temperature value depend on the thermodynamic model which is applied to derive the temperature. The measured temperature is usually determined relative to the internal energy or the shape of the spectrum based on the model. If the assumed model is different, the derived value of temperature differs. For example, if the temperature is determined by the total internal energy, Eq. \ref{eq:Uqm} and Eq. \ref{eq:Ufvc} needs to be equal. If the model which is based on quantum mechanics is used,
\begin{equation}
T_m \approx T_{qm}\approx \sqrt[4]{0.41}T_{FV}=0.80T_{FV}
\label{eq:T}
\end{equation}
where $T_m$ is the measured temperature, $T_{qm}$ is $T$ in Eq. \ref{eq:Uqm} and $T_{FV}$ is $T$ in Eq. \ref{eq:Ufvc}.

In general, when we compare different models, we should not expect derived $T_{qm}$ and $T_{FV}$ to be the same value for the same experimental condition. In order to compare each model to the measurement, we need to look into details on how the measured temperature was determined by experiment and use the temperature definition for the each model ($T_{qm}$ is T in Eq. \ref{eq:Bqm} and $T_{FV}$ is T in Eq. \ref{eq:Bfvc}) to convert to each other. $T$ which is derived from the correct model would retain the fundamentally meaningful parameter as temperature. The other $T$ would be just a fitting parameter to an incorrect formula.

We can also derive Wien's displacement law from TVT with filled vacuum; $\lambda^{FV}_{max}=b^{FV}/T$ where $b^{FV}=3.77\times10^{-3}$ mK $\approx 1.30b$. Again, there is nothing to worry about the factor, 1.30. Since the assumed models are different, the derived temperatures from measurement of $\lambda_{max}$ would give $T_{FV} \approx 1.30 T_{qm}$.

\begin{figure}[h]
\includegraphics[trim={0cm  0cm  1cm 0cm},clip,width=0.5\textwidth]{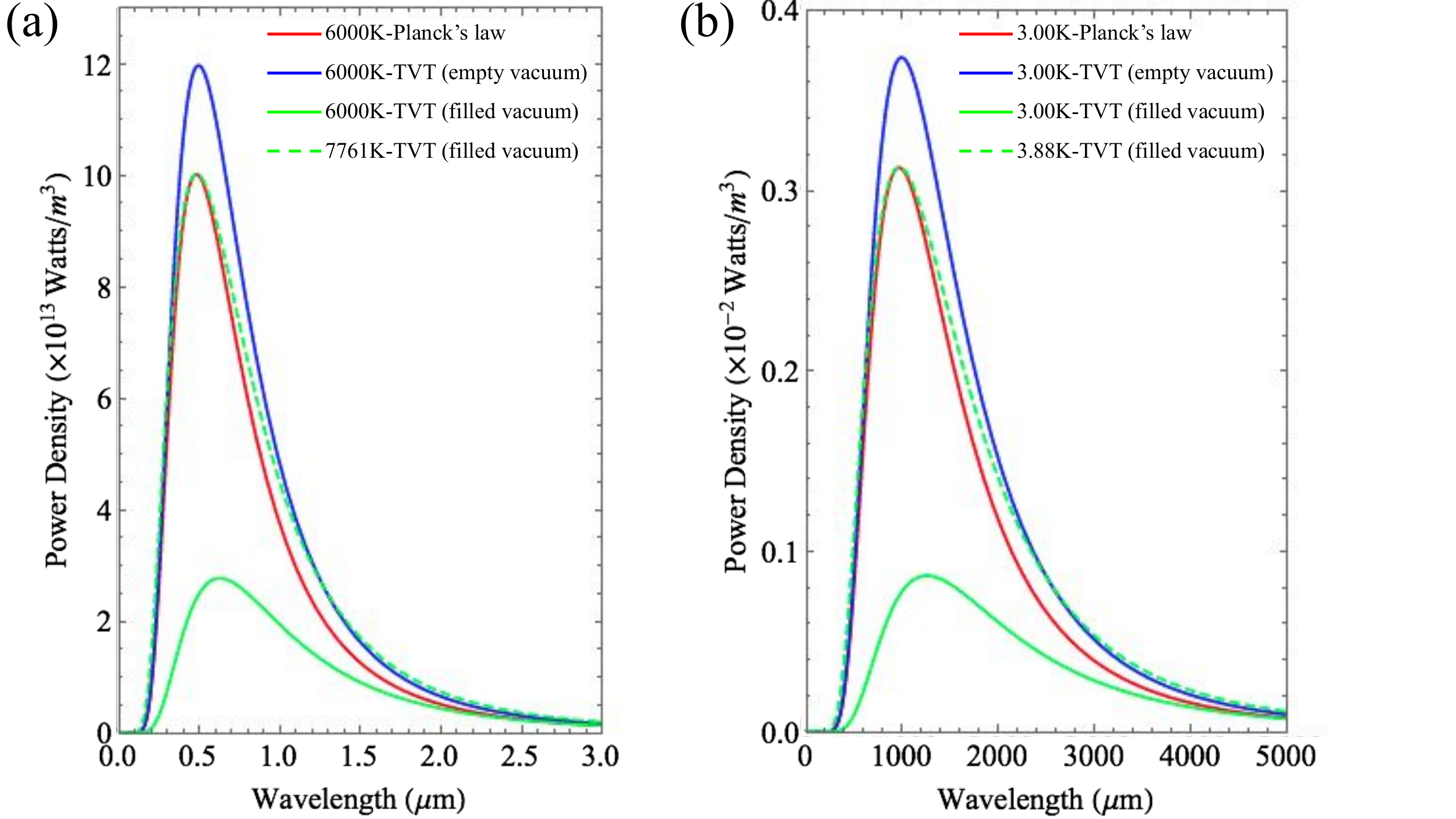}
\caption{The radiated power density per wave length, $S_\lambda(\lambda, T)$ is plotted for quantum mechanics (red), TVT with empty vacuum (blue) and TVT with filled vacuum (green) respectively in (a) at 6000 K (7761 K also for TVT with filled vacuum in dotted green curve) and (b) at 3.00 K (3.88 K also for TVT with filled vacuum in dotted green curve).}
\label{fig:BBschematicF}
\end{figure}

In Fig. \ref{fig:BBschematicF}, the radiated power density per wavelength, $S_\lambda^{qm}(\lambda, T)$, $S_\lambda^{V}(\lambda, T)$ and $S_\lambda^{FV}(\lambda, T)$ are plotted for quantum mechanics, TVT with empty vacuum and TVT with filled vacuum respectively where $S_\lambda^{FV}(\lambda, T)=\frac{2\pi c}{\lambda^4}(\frac{hc}{\lambda}+k_B T)/((hc/\lambda k_B T)e^{hc/\lambda k_B T}+1)$. As expected, at the same T value, the power density has larger values for TVT with empty vacuum than for quantum mechanics since more dissipative energy states (the orange part Fig. \ref{fig:BB} (b)) are allowed for TVT with empty vacuum than for quantum mechanics (the orange lines Fig. \ref{fig:BB} (a)), and it has smaller values for TVT with filled vacuum than for quantum mechanics since there are more ground states (the purple part Fig. \ref{fig:BB} (b)) filled in the cavity modes for TVT with filled vacuum than for quantum mechanics (the purple line Fig. \ref{fig:BB} (a)). While one (or none) of these three models should describe the reality more accurately than others, it would be hard to determine it by blackbody radiation experiments alone since their temperature dependence and spectral shapes are very similar.

\begin{figure}[h]
\includegraphics[trim={1.5cm  0cm  1cm 0cm},clip,width=0.4\textwidth]{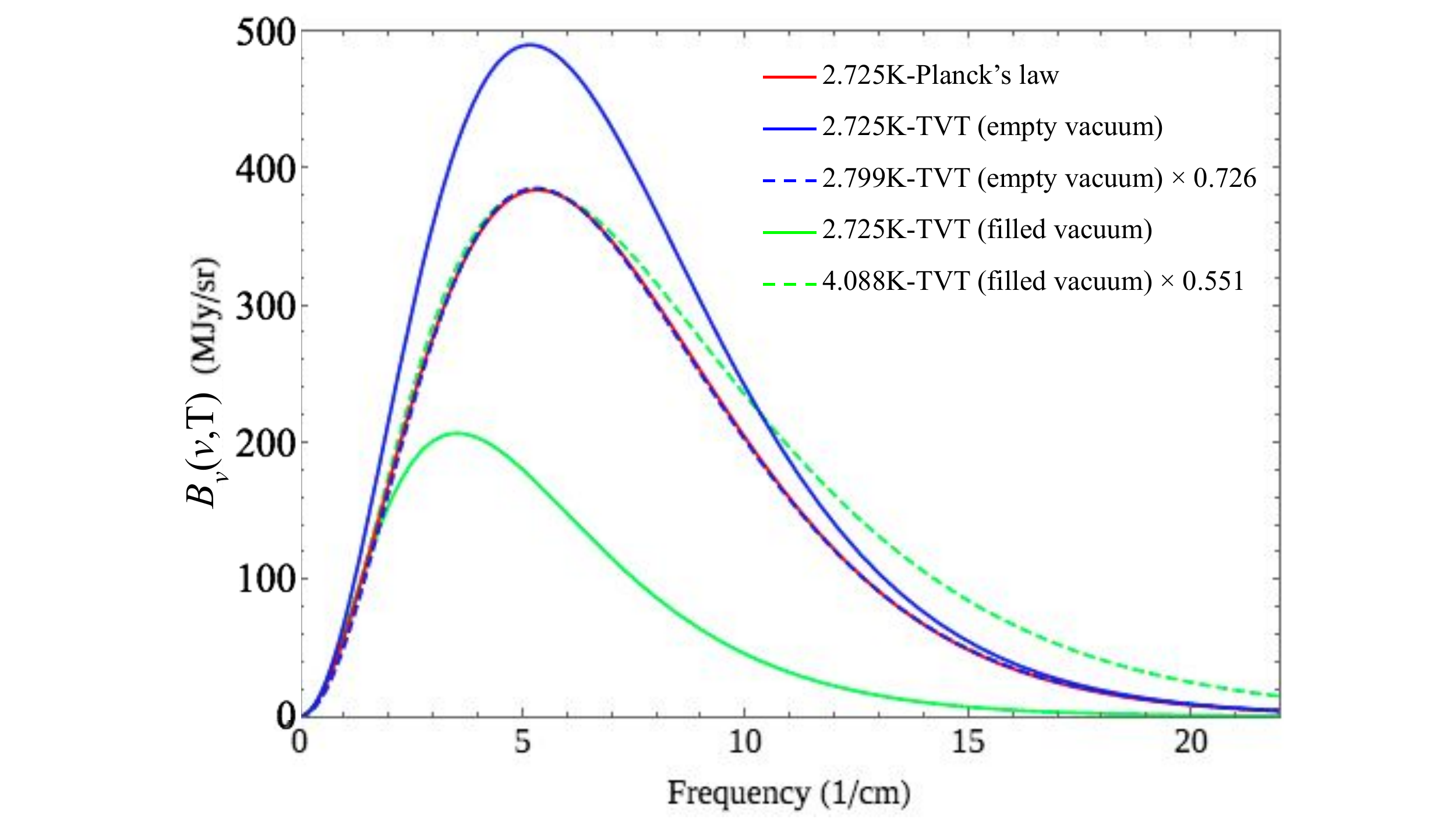}
\caption{The spectral radiance, $B_\nu(\nu,T=2.725K)$ MJy/sr is plotted for Planck's law (solid red), TVT in empty vacuum with Eq. \ref{eq:Bvc} (solid blue) and TVT in filled vacuum with Eq. \ref{eq:Bfvc} (solid green) as a function of frequency (please note that in Fig. \ref{fig:BBschematicF}, $S_\lambda(\lambda, T)$ is plotted in a function of wavelength). The dotted blue curve is for TVT in empty vacuum with relatively small adjustments in parameters and almost overlaps the red curve. The dotted green curve is for TVT in filled vacuum and would not fit the red curve even with relatively large adjustments in parameters. This might show that CMB was emitted in empty vacuum rather than filled vacuum. This could be an evidence of an expansive universe in the early stage.}
\label{fig:CMB}
\end{figure}

In Fig. \ref{fig:CMB}, the spectral radiance, $B_\nu^{qm}(\nu,T)$ is plotted in MJy/sr for Planck's law at 2.725 K (solid red). This red curve is known to perfectly fit the experimental observation of cosmic microwave background radiation (CMB) \cite{CMB}. $B_\nu^{V}(\nu,T)$ MJy/sr for TVT with empty vacuum in Eq. \ref{eq:Bvc} at 2.725 K is also plotted in solid blue curve. A dotted blue curve represents $0.726 \times B_\nu^{V}(\nu,T=2.799K)$ MJy/sr which is at the slightly higher temperate than 2.725K with the amplitude scaled by 0.726. This dotted blue curve almost completely overlaps the red solid curve. Therefore, most likely, TVT with empty vacuum would fit the experimental observation of CMB as well as planck's law. The solid and dotted green curves are for TVT with filled vacuum in Eq. \ref{eq:Bfvc}.

\section{\label{sec:Debye model}Debye model}

In this section, the Debye model will be reviewed for the case of TVT with filled vacuum. Using the Bose-Einstain distribution in Eq. \ref{eq:FBE}, the internal energy with the Debye model is expressed as
\begin{multline}
\label{eq:UqmD}
U^{qm}=9N k_B T(T/T_D)^3 \int^{T_D/T}_0 x^3 f_{BE}(x)dx
\end{multline}
where $T_D$ is the Debye temperature, and $N$ is the number of atoms in the solid \cite{Debye}. The heat capacity for quantum mechanics is:
\begin{multline}
\label{eq:CVqmD}
C_V^{qm}=\frac{\partial U^{qm}}{\partial T}
=9N k_B (T/T_D)^3 \int^{T_D/T}_0 \frac{x^4 e^x}{(e^x -1)^2}dx.
\end{multline}

\begin{figure}[h]
\includegraphics[trim={0cm  0cm  0cm 0cm},clip,width=0.48\textwidth]{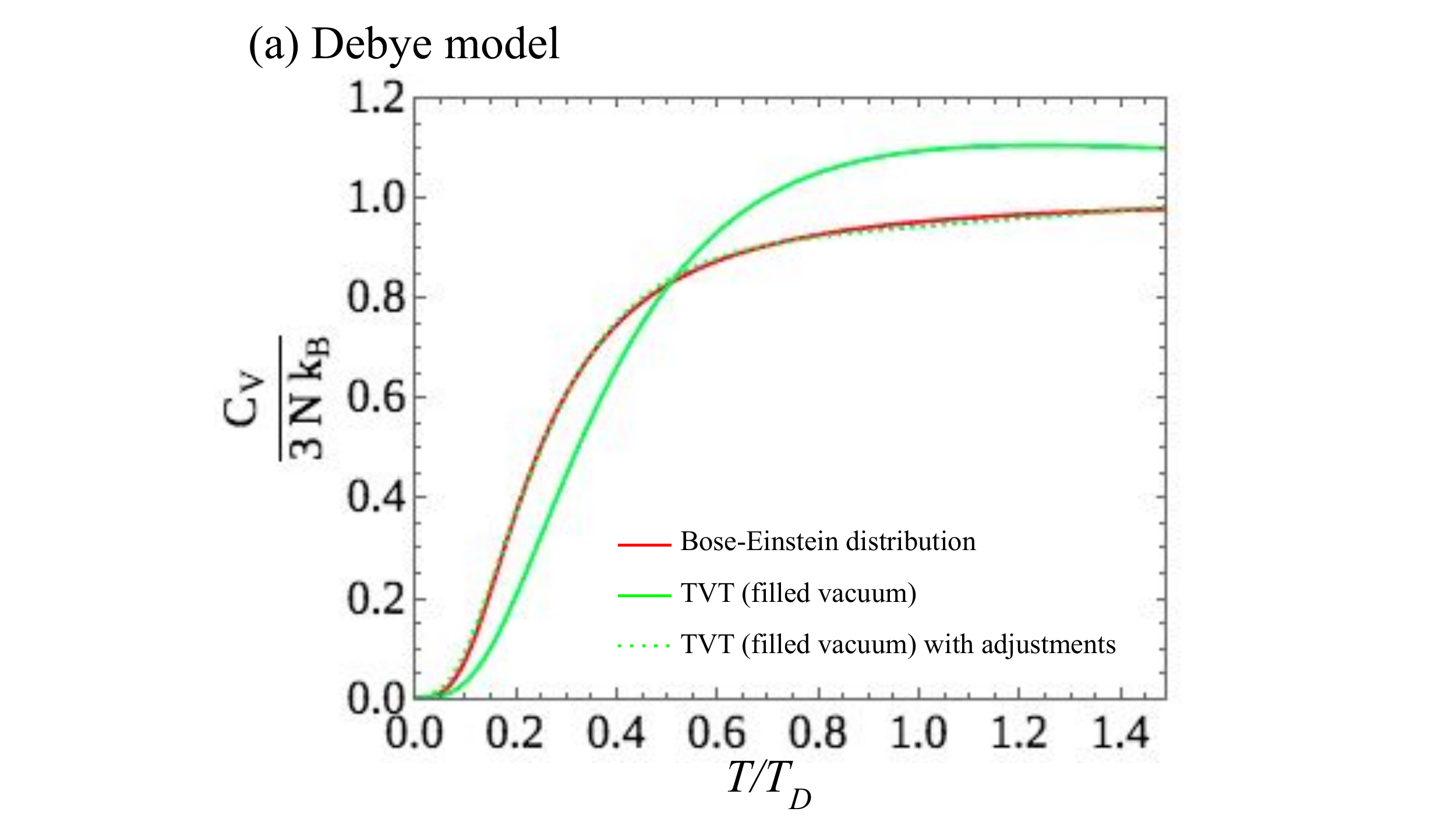}
\includegraphics[trim={0cm  0cm  0cm 0cm},clip,width=0.48\textwidth]{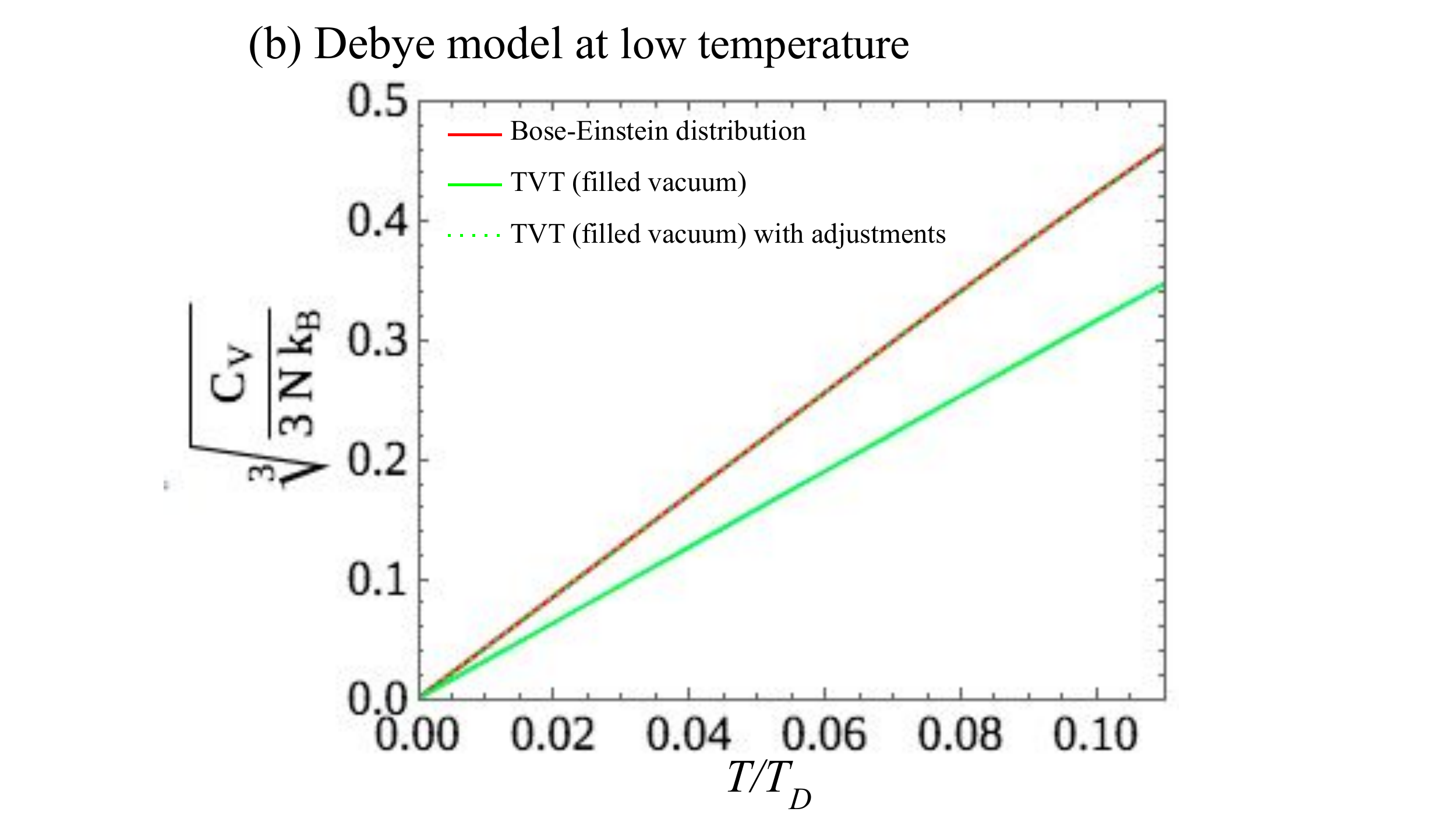}
\caption{\textbf{(a)} Eq. \ref{eq:CVqmD} is plotted for $C_V^{qm}/3Nk_B$ (solid red curve), and Eq. \ref{eq:CVfvc} is plotted for $C_V^{FV}/3Nk_B$ (solid green curve) as a function of $T/T_D$. As mentioned before, in the different models, derived values of $T$ for different models are different for the same experimental conditions. This makes it difficult to simply compare these two equations. At relatively high temperatures, the simple conversion such as Eq. \ref{eq:T} can not be used since many thermometric materials have their Debye temperature within the temperature range. Relatively justifiable three free parameters, which are often used by experimental analyses, are added to the solid green curve to fit the solid red curve to show that it would be hard to distinguish the two models by experiments. $C_V^{FV}/3Nk_B=a (3(T/bT_D)^3 \int^{b T_D/T}_0 \frac{x^3 e^x (x^2 +2x+2)+x^2}{(x e^x +1)^2}dx+c T/bT_D)$ is plotted in the dotted green curve. $a=0.74$ is added for scaling for detector sensitivity, $b=0.65$ is added because $T_D$ should be a free parameter and could be different for a different model, and $c=0.11$ is added because conducting electron heat capacity, which has linear dependency to $T$, is often subtracted. \textbf{(b)} Eq. \ref{eq:CVqmD} is plotted for $\sqrt[3]{C_V^{qm}/3Nk_B}$ (solid red curve), and Eq. \ref{eq:CVfvc} is plotted for $\sqrt[3]{C_V^{FV}/3Nk_B}$ (solid green curve) in a function of $T/T_D$ at low temperature. $\sqrt[3]{C_V^{FV}/3Nk_B}$ was plotted in a function of $T/0.80 T_D$ from Eq. \ref{eq:T} (dotted green curve).}
\label{fig:Debye}
\end{figure}

For TVT with filled vacuum, the solid is at the thermal equilibrium with the environment including the vacuum. We can safely assume that the same distribution function for TVT with filled vacuum in Eq. \ref{eq:Ffvc} can be used for atomic lattice vibrations. Therefore, 
\begin{multline}
\label{eq:UfvcD}
U^{FV}=9N k_B T(T/T_D)^3 \int^{T_D/T}_0 x^3 f_{FV}(x)dx.
\end{multline}
The heat capacity for TVT for filled vacuum is:
\begin{multline}
\label{eq:CVfvc}
C_V^{FV}=\frac{\partial U^{FV}}{\partial T}\\
=9N k_B (T/T_D)^3 \int^{T_D/T}_0 \frac{x^3 e^x (x^2 +2x+2)+x^2}{(x e^x +1)^2} dx,
\end{multline}
which is plotted in solid green curves in Fig. \ref{fig:Debye}.

At the low temperature limit of $T \ll T_D$, Eq. \ref{eq:CVfvc} becomes $C_V^{FV}=10.6 \times 9N k_B (T/T_D)^3$ while $C_V^{qm}=26.0 \times 9N k_B (T/T_D)^3$. TVT has the correct $T^3$ dependence which is observed by experiments. Again, in order to compare the temperature dependence with experiments, the temperature needs to be extracted with the same model to be consistent. The different factors can be explained by the different definitions of temperature, which can be converted by Eq. \ref{eq:T} (see Fig. \ref{fig:Debye} (b)). At the high temperature limit of $T \gg T_D$ for Eq. \ref{eq:CVfvc}, $C_V^{FV}\rightarrow 3N k_B$ to the Dulong–Petit law.

In Fig. \ref{fig:Debye}, the dotted greens curves shows examples of good fits, after justifiable adjustments, to the red curves which are derived with Bose-Einstein distribution. In any case, there seem to be no apparent discrepancies with past heat capacity experiments of lattice vibration with the model of TVT (filled vacuum). It would be very hard and not definitive to determine which of Bose-Einstein distribution or TVT (filled vacuum) distribution provides a better fit to experimental data. It would be a good idea to examine other experiments for that purpose. 

\section{\label{sec:Massive particles}Massive particles}
In previous sections, we considered massless energy distributions such as electromagnetic fields and lattice vibrations. In this section, we will examine a distribution for massive particles. Unlike the massless energy distribution, the vacuum fluctuations (kinetic energy less than the threshold) of massive particles are directly observable and measurable for cases such as pressure, current and spectroscopic absorption measurements since the particles themselves would not disappear even when their kinetic energy is immersed in the vacuum energy. Therefore, for the case of massive particles, the vacuum energy contribution needs to be added in the summation of the internal energy. Most of the cases, massive particles would have interaction with the vacuum energy as sound waves. The spectral energy distribution for massive particles with frequency $\nu$ at temperature $T$ is
\begin{multline}
B_\nu^{MV}(\nu,T)\\
=\lim_{T_V\to\infty}\widetilde{g}(\nu)\ \frac{\int_{0}^{h\nu} E\,e^{-\frac{E}{k_B T_V}}dE+\int_{h\nu}^\infty E\,e^{-\frac{E}{k_B T}}dE}{\int_{0}^{h\nu} e^{-\frac{E}{k_B T_V}}dE+\int_{h\nu}^\infty e^{-\frac{E}{k_B T}}dE}\\
=\widetilde{g}(\nu)\ h\nu\ f_{MV}(h\nu/k_BT)
\label{eq:Bmvc}
\end{multline}
where
\begin{equation}
f_{MV}(x)=\frac{\frac{1}{2}+\frac{1}{x}}{xe^{x}+1}+\frac{1}{2}.
\label{eq:Fmvc}
\end{equation}

\begin{figure}[h]
 \includegraphics[trim={0.15cm  1.1cm  7.13cm 0cm}, clip, width=0.49\textwidth]{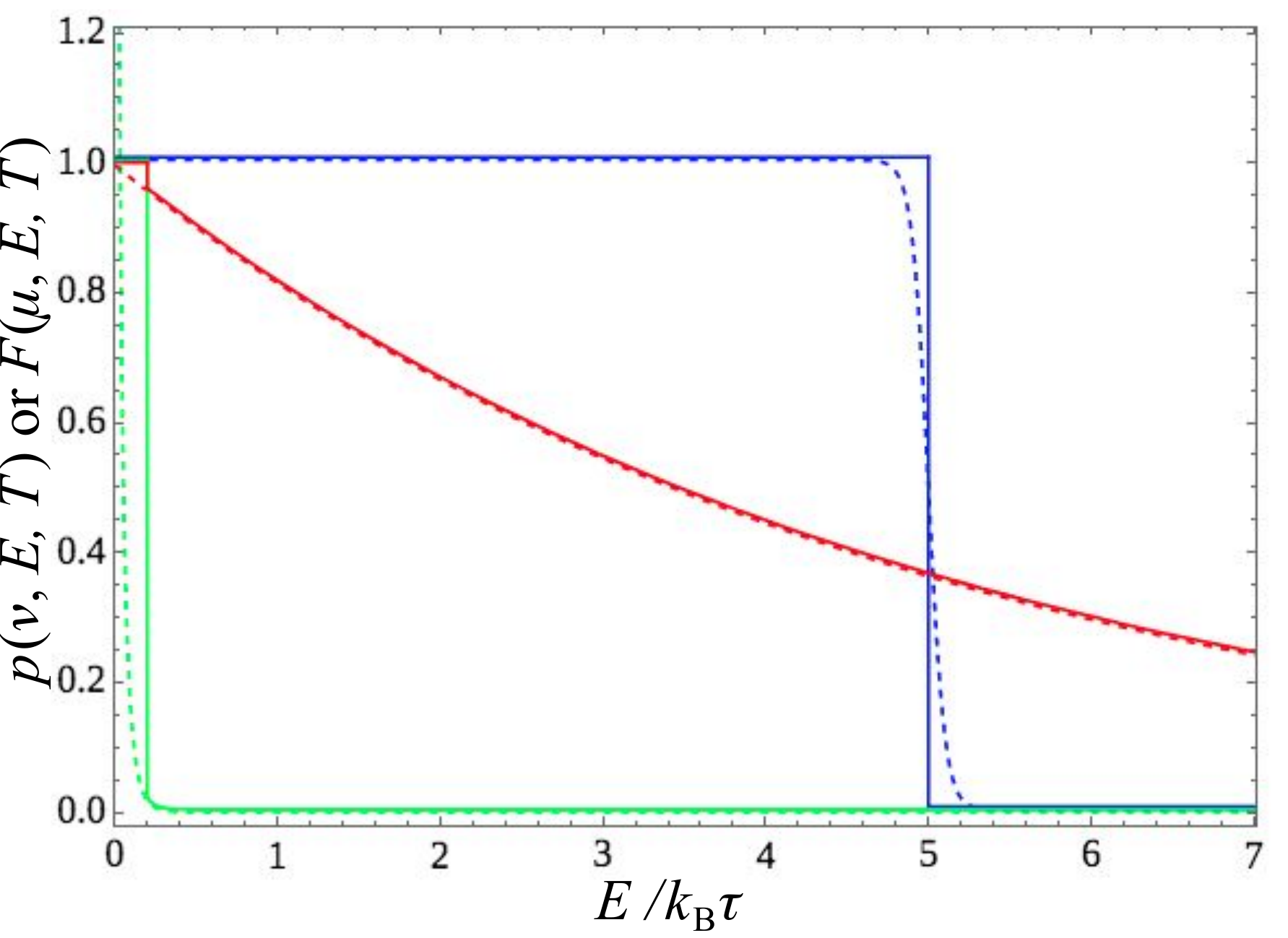}
  \includegraphics[trim={0cm  11.2cm 6.5cm 0cm}, clip, width=0.49\textwidth]{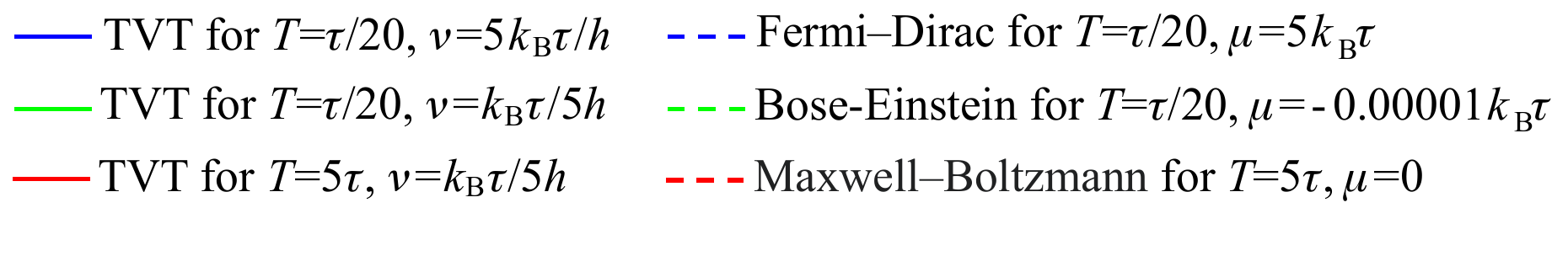}
\caption{Eq. \ref{eq:p} is plotted as a function of $E/k_B \tau$ for two temperatures $T=\tau/20$ \& $5\tau$, two frequency $\nu=k_B \tau/5h$ \& $5k_B \tau/h$ in three combinations in solid lines. Appropriate conventional distributions $F(\mu, E, T)$ are applied to represent $p(\nu, E, T)$ in TVT for these three cases. Fermi-Dirac distribution $F(\mu, E, T)=1/(e^{(E-\mu)/k_B T}+1)$ (dashed blue) represents $p(\nu, E, T)$ for case of high $\nu$ and low $T$ (solid blue). Bose-Einstein distribution $F(\mu, E, T)=1/(e^{(E-\mu)/k_B T}-1)$ (dashed green), represents $p(\nu, E, T)$ for case of low $\nu$ and low $T$ (solid green). Maxwell-Boltzmann distribution $F(\mu, E, T)=e^{-(E-\mu)/k_B T}$ (dashed red) represents  $p(\nu, E, T)$ for case of low $\nu$ and high $T$ (solid red). The value of $\mu$ is adjusted so that $\int_0^\infty F(\mu, E, T)dE=\int_0^\infty p(\nu, E, T)dE$.}
\label{fig:p}
\end{figure}

$B_\nu^{MV}(\nu,T)$ is different from Eq. \ref{eq:Bfvc} only by the added first term in the numerator \cite{CV}, and the density of modes for the frequency $\nu$ is replaced with $\widetilde{g}(\nu)$. $f_{MV}(h\nu/k_BT)$ is the distribution function for massive particles in TVT with filled vacuum, and it is the average fraction of energy relative to $h\nu$ per mode for the frequency $\nu$ at  temperature T. It is easy to see the average energy for each mode, $h\nu f_{MV}(h\nu/k_BT) \rightarrow k_BT$ when $k_B T \gg h\nu$ (classical limit), and $h\nu f_{MV}(h\nu/k_BT) \rightarrow h\nu/2$ when $h\nu \gg k_B T$ (quantum limit). Eq. \ref{eq:Fmvc} is plotted in Fig. \ref{fig:f} in dotted purple. 

Now, let us review the meaning of Eq. \ref{eq:Bmvc} and the role of $\widetilde{g}(\nu)$ for massive particles. For TVT, we only use the Boltzmann factor of the probability $p(\nu, E, T)$ for the mode of frequency $\nu$ being excited to the energy $E$ at the temperature $T$ in the filled vacuum. 
\begin{multline}
p(\nu, E, T)
=1+ H(E-h\nu)(e^{-\frac{E}{k_B T}}-1)
\label{eq:p}
    \end{multline}
where $H(x)$ is the Heaviside step function. Eq. \ref{eq:Bmvc} is obtained from the expectation value of E with $p(\nu, E, T)$ weighted by the density of modes $\widetilde{g}(\nu)$. $B_\nu^{MV}(\nu,T)=\widetilde{g}(\nu)\int_{0}^\infty E p(\nu, E, T)dE/\int_{0}^\infty p(\nu, E, T)dE$.  $\widetilde{g}(\nu)$ highly depends on the system such as dimensions (3D, 2D or 1D), particle interactions (solid, liquid or gas), kinds of particles (monoatomic, charge, spin, etc.) and kinds of potential (free in a box, harmonic potential or ion lattices).

For conventional distributions for massive particles, there is no labeling for $\nu$ nor $\widetilde{g}(\nu)$. Instead, they have the density of energy $g(E)$. In order to compare the two methods, we need to integrate $\widetilde{g}(\nu)p(\nu, E,T)$ over $\nu$ and show $\int_0^{\infty}\widetilde{g}(\nu)p(\nu, E, T)d\nu \approx g(E)F(\mu,E,T)$ where $F(\mu, E,T)$ is the conventional distributions. In Fig. \ref{fig:p}, Eq. \ref{eq:p} is plotted as a function of $E/K_B \tau$ for different $\nu$ and $T$. It can be easily inferred that fermions should have an optical sound-like distribution of $\widetilde{g}(\nu)$ where $\nu$ is distributed at relatively high average value such as a Gaussian distribution. This is a reasonable requirement since many fermions have a charge and also often in an ion-lattice potential. Bosons should have an acoustic sound-like distribution of $\widetilde{g}(\nu)$ with a cut off value of $\nu_{max}$. The existence of the filled vacuum alone seems to easily explain many quantum phenomena for both fermions and bosons in condensed matter physics at relatively low temperature without using quantum statistics. Each system is very unique, and the details for each phenomenon should be left to the experts in each field.

The average number of massive particles with energy $E$ at $T$ can be expressed with the conventional distributions as
$\widetilde{N}(E,T)=g(E)F(\mu,E,T)$ where $\mu$ is defined by $N=\int_0^\infty g(E)F(\mu,E,T)dE$ and $N$ is the total number of the particles. For TVT,
\begin{multline}
\widetilde{N}(E,T)
=N\frac{\int_0^\infty \widetilde{g}(\nu)p(\nu,E,T)d\nu}{\int_0^\infty\int_0^\infty \widetilde{g}(\nu)p(\nu,E,T)d\nu dE}.
\end{multline}

\section{\label{sec:Thermodynahttps://www.overleaf.com/project/611ab1e64f989825fb5f75e0mics}Thermodynamics}
\begin{figure}[h]
\includegraphics[trim={0cm  1.5cm 8cm 1cm},clip,width=0.35\textwidth]{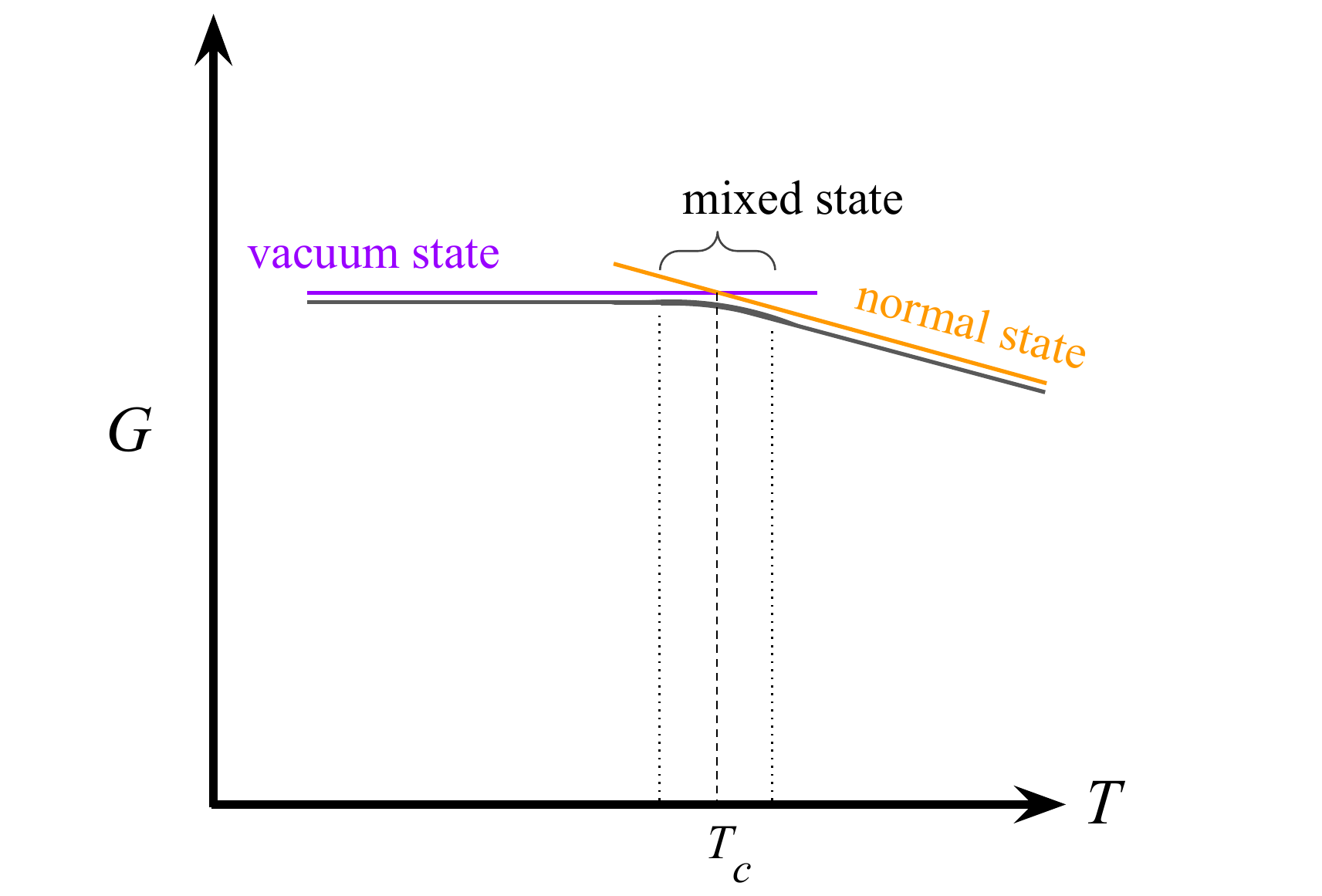}
\caption{The Gibbs free energy for the vacuum sate $G_V(P,T_V)$, the mixed state $G_M(P,T,T_V)$ and the normal state, $G_N(P,T)$ are plotted as a function of $T$ though the phase transition. The vacuum state doesn't have $T$ dependence. Therefore{\color{red},} $G_V(P,T_V)$ is a constant value. In order to have this phase transition, $T_VS_V$ needs to be relatively large so that $G_V(P,T_V)<G_N(P,T)$ at low temperature. $T_V$ is a very large value. However, $S_V$ is a very small value because of the threshold energy unless high frequency states are available. The high frequency modes have larger energy threshold, and the density of modes grows with $\nu^2$ for three dimension. As $T$ is reduced, massive particles become motionless relative to each other and can be locked into position with deviations of small amplitude but high frequency vibration at the ordered positions. In order to achieve higher frequency modes, the particle mass needs to be small for a higher speed of sound, the nearest distance needs to be small (the density of particles needs to be large) for the small minimum wave length, and the pressure needs to be large (the inter-particle force needs to be large) for the higher speed of sound. These conditions coincide the conditions for phase transitions of ordered states in quantum mechanics.}
\label{fig:Gibbs}
\end{figure}

According to TVT, electrons in a metal may not be free in an ion lattice contrary to the commonly accepted view. Instead, they are frozen into the resonant mode of $\nu_R$. (An electron has a charge, and its electromagnetic potential is relatively strong. It also has a small mass. Those factors make the resonant frequency relatively high.) This means that through vacuum interaction, electrons absorb small amounts of energy due to vibrations of $\nu_R$ ($E < h\nu_R$) at the resonant position. In other words, it condenses into an ordered state. Once it receives the vacuum energy, it will not be able to let go easily when at equilibrium with the vacuum since the vacuum is filled with such energy. It is trapped in the state. It will need higher energy i.e. high temperature, $T$ for the normal (dissipative) state to remove an electron away from the resonant position (phase transition to the normal state). There could be many metastable resonant states in the vacuum states with potential wells of various depths which can be reduced by parameters such as electric \& magnetic fields and pressure.  

The Gibbs free energy is given by:
\begin{multline}
\begin{cases}
G_V(P,T_V)=U+PV-T_V S_V\\
G_M(P,T, T_V)=U+PV-T_V S_V-T S_N\\
G_N(P,T)=U+PV-T S_N
\end{cases}
\end{multline}
where $G_V(P,T_V)$ is for the vacuum state, $G_M(P,T,T_V)$ is for the mixed state and $G_N(P,T)$ is for the normal state. They are plotted in Fig. \ref{fig:Gibbs}. $S_V$ is the entropy for the vacuum state, $S_N$ is the entropy for the normal states, $U$ is the internal energy, $P$ is the pressure and $V$ is the volume. In this case, $T_V$ is a finite large value, and $T_V$ should not be approximated by taking the limit as $T_V\rightarrow \infty$. Even though there is a phase transition at $T=0$, it is also a thermodynamic phase transition since $G_V(P,T_V)$ still has a temperature term in it. $T_V$ has a very particular energy distribution with a frequency-dependent threshold and may be controllable with improved technology.

Realizing the existence of the vacuum energy and introducing the parameters like, $T_V$ and $S_V$ solves many paradoxes in thermodynamics. Thermodynamics with TVT will provide a more consistent and versatile theory which can predict the future of the universe. Indeed, the universe is all about the dance between $S_V$ and $S_N$. This mechanism is embedded in the threshold condition of vacuum. The energy above the threshold tends to dissipate and be disordered toward the larger value for $S_N$. The energy below the threshold tends to be condensed and ordered toward the larger value for $S_V$. More details will be discussed in other article \cite{Yoko2}.

\section{\label{sec:concludion}Conclusion}

Theory of Vacuum Texture (TVT) is a deterministic theory with a local hidden variable. The microscopic theory of vacuum is formulated with the single postulate, the threshold condition. TVT agrees well with experimental observation of black body radiation, uncertainty principle and quantum statistics. TVT may explain the experimental observation without quantum mechanics and the ``spooky'' phenomena associated with it. We showed, TVT and quantum mechanics give similar results. More experiments and data analysis are needed to distinguish the two \cite{Yoko1}. We must consider other phenomenon, such as gravitation, in order to determine if quantum mechanics or TVT are viable theories \cite{Yoko2}.

\begin{acknowledgments}
This work was supported by the M. Hildred Blewett Fellowship of the American Physical Society, www.aps.org.
\end{acknowledgments}


\begin{thebibliography}{99}
\bibitem{alphaT}Please note that $T_V=\alpha T$ needs to be a fixed value. Therefore, Eq. \ref{eq:Bpvc} and \ref{eq:Fpvc} are good for a fixed T and can not be used to prove T dependence.
\bibitem{CV}Please note that for heat capacity measurements, the first term in the numerator should not be added since the vacuum energy is not thermally isolated from the environments, and that heat capacity is a derivative of $T$ (and not $T_V$).
\end{thebibliography}
\end{document}